\documentclass[11pt]{article}
\usepackage{amsmath, amssymb, amscd, amsthm, amsfonts}
\usepackage{graphicx}
\usepackage[numbers,sort&compress]{natbib}
\usepackage{hyperref}
\usepackage[affil-it]{authblk}
\usepackage[numbers,sort&compress]{natbib}
\usepackage{booktabs}
\usepackage{bm}
\usepackage{xfrac}
\usepackage{nicefrac}
\usepackage{caption}
\usepackage[T1]{fontenc}
\usepackage[utf8]{inputenc}
\DeclareMathOperator*{\argmin}{arg\,min}

\title{Comparing Variable Selection and Model
Averaging Methods for Logistic Regression}

\author[1]{Nikola Sekulovski}
\author[1]{Franti\v{s}ek Barto\v{s}}
\author[1]{Don van den Bergh}
\author[1]{Giuseppe Arena}
\author[1]{Henrik R. Godmann}
\author[1]{Vipasha Goyal}
\author[1]{Julius M. Pfadt}
\author[1]{Maarten Marsman}
\author[2]{Adrian E. Raftery}

\affil[1]{Department of Psychology, University of Amsterdam, The Netherlands}
\affil[2]{Department of Statistics, University of Washington, USA}

\begin{document}

\maketitle

\begin{abstract}
Model uncertainty is a central challenge in statistical models for binary outcomes such as logistic regression, arising when it is unclear which predictors should be included in the model. Many methods have been proposed to address this issue for logistic regression, but their relative performance under realistic conditions remains poorly understood. We therefore conducted a preregistered, simulation-based comparison of 28 established methods for variable selection and inference under model uncertainty, using 11 empirical datasets spanning a range of sample sizes and number of predictors, in cases both with and without separation. We found that Bayesian model averaging (BMA) methods based on $g$–priors, particularly $g = \max(n, p^2)$, show the strongest overall performance when separation is absent. When separation occurs, penalized likelihood approaches, especially the LASSO, provide the most stable results, while BMA with the local empirical Bayes (EB-local) prior is competitive in both situations. These findings offer practical guidance for applied researchers on how to effectively address model uncertainty in logistic regression in modern empirical and machine learning research.
\end{abstract}

\section*{Introduction}
Model uncertainty is a central challenge in statistical inference, especially in regression settings where it is unclear in advance which predictors should be included in the model. In this context, each possible subset of the predictors can constitute a model, and even a modest number of predictors can give rise to a large number of possible models. Many methods have been proposed for addressing this problem, including both Bayesian and frequentist approaches. However, even though regression models are among the most widely used tools in applied statistics, the relative performance of these methods in realistic empirical settings had not been systematically evaluated. Porwal and Raftery \cite{PorwalRaftery_2022} addressed this by conducting an extensive comparison of 21 commonly used methods for the linear regression model. They found that performance varied widely, and that three adaptive Bayesian model averaging approaches, which adapt to the analysis design and account for model uncertainty, performed best overall. Their results provide practical guidance for applied researchers and a benchmark for further methodological development.

Similar challenges arise when the outcome is binary and logistic regression is used. Logistic regression \citep{Berkson1944,Cox1958,HosmerLemeshow2013} is a standard tool for modeling binary outcomes and is widely applied in fields such as epidemiology, social science, and machine learning. Here, we consider the logistic regression model
\[
\text{logit}(P(Y_i = 1 \mid \mathbf{x}_i)) = \beta_0 + \sum_{j=1}^p \beta_j x_{i,j}, \quad i = 1, \dots, n,
\]
where $Y_i \in \{0,1\}$ is a binary outcome and $\mathbf{x}_i = (x_{i,1}, \dots, x_{i,p})^\top \in \mathbb{R}^p$ is the vector of candidate predictors for observation $i$. The coefficients $\beta_0, \dots, \beta_p$ describe the log-odds relationship between the predictors and the outcome.

As in the linear case, a major challenge is model uncertainty, which arises when there is a set of candidate predictors and it is unclear which ones to include in the model. In addition to uncertainty about variable inclusion, logistic regression introduces specific estimation difficulties. One such difficulty is separation, which occurs when a linear combination of predictors perfectly classifies the outcome. In these cases, unique maximum likelihood estimates may not exist \citep{albert1984existence}, leading to numerical instability and invalid inference. Separation is particularly common in small samples and high-dimensional settings, where it can arise even when the true signal is weak.

While there is broad agreement on how to conduct inference when the model structure is known and fixed, it often is not in practice. In many applications, the set of relevant covariates is not fully known in advance and must be determined at least partly from the data. This introduces an additional layer of uncertainty that affects both estimation and inference. A wide range of methods have been proposed to address model uncertainty and estimation instability in logistic regression, but their comparative performance under realistic conditions remains unclear. Our goal is to address this gap by systematically evaluating 28 established methods for statistical inference in logistic regression under model uncertainty.

Early applications of logistic regression typically relied on subject-matter expertise to determine which covariates to include in the model. This approach reflected the scientific priorities of the time, which emphasized interpretability and theoretical grounding, and was shaped by the computational constraints of early statistical analysis. Arguments and guidelines for doing this are described by \cite{HosmerLemeshow2013,Bursac2008,Stoltzfus2011}. 

As logistic regression became more widely adopted and software implementations became more accessible, some analysts turned to a more inclusive strategy, fitting models with all available covariates. While this removed the need for subjective selection, it introduced new risks, including overfitting and degraded predictive performance. Stepwise selection procedures emerged as a practical alternative and gained popularity throughout the 1980s and 1990s \citep{HosmerLemeshow2013}. These algorithms add or remove covariates based on statistical criteria such as likelihood ratio tests or information criteria. Even though stepwise methods have well-documented shortcomings \citep{Derksen1992, miller2002subset,Freedman1983,Austin2004}, especially when there is separation \citep{heinze2006comparative, kotani2025goodness}, they remain widely used to this day due to their simplicity and availability in standard software packages.

In response to these limitations, the past three decades have seen the development of more principled and systematic approaches to variable selection under model uncertainty in logistic regression. Most of these methods fall into two broad categories: Bayesian approaches and penalized likelihood techniques.

Bayesian model averaging (BMA) is a widely studied framework for addressing model uncertainty in statistical inference. Instead of selecting a single model, BMA combines inferences across all plausible models, weighting each according to its posterior probability given the observed data. This approach offers a principled way to account for model uncertainty and has been shown to improve estimation and prediction in a variety of settings. Following its early development \citep{Leamer1978,Raftery1988,George1993,Madigan1994}, it has since been extensively applied to generalized linear models, including logistic regression \citep{Tuchler01032008,li_clyde_2018,ChenEtAl_1999,Wagner_2012,Raftery_1996_approx,Porwal_2024}. Several comprehensive reviews summarize the theoretical foundations and practical applications of BMA in regression \citep{kass1995bayes,Hoeting1999,fernandez2001benchmark,Wasserman2000,Clyde2004,Fragoso2018,Forte2018,Kaplan_2021}. All of the Bayesian methods included in our comparison use BMA.

BMA has some good theoretical properties in general \citep{raftery2003discussion}. BMA point estimators and predictions minimize mean squared error (MSE), and BMA estimation and prediction intervals are calibrated. BMA predictive distributions have optimal performance in the log-score sense \citep{Madigan1994}. These properties hold on average over the prior distribution, extending similar results for Bayesian estimation \citep{RubinSchenker1986}, but the results are somewhat robust to this assumption \citep{Mattei2020}. When the prior distribution is used as the distribution of parameter values over which performance is averaged, it has been  referred to as the world distribution \citep{Jeffreys1961}, the practical distribution \citep{raftery2003discussion}, or the effect-size distribution \citep{park2010estimation}. Analysis using this concept has been said to follow the ``empirical frequentist principle'' \citep{Berger2021}.

The implementation of BMA involves several choices by the user, including the prior distribution of the model parameters under each model, and the prior model probabilities. Also, the number of candidate models can be too large for them all to be feasibly evaluated. For example the number of possible subsets of $p$ regression variables is $2^p$; for $p$ much beyond 25 or 30 this can be computationally prohibitive. Thus the choice of analytic or computational approximations must also be made. Together these choices lead to many possible implementations of BMA.

Bayesian model averaging for logistic regression typically relies on adaptations of prior distributions originally developed for linear regression models. In particular, variants of Zellner’s $g$-prior \citep{zellner1986assessing} and its extensions, including mixtures of $g$-priors, empirical Bayes adaptations \citep{clyde2000flexible, george2000calibration}, and fully Bayesian formulations such as the hyper-$g$ prior \citep{liang2008mixtures}, have been adapted for use in generalized linear models \citep{Tuchler01032008, li_clyde_2018}. These priors are widely implemented and support both variable selection and model averaging, although their theoretical justification is strongest under the Gaussian likelihood. In practice, they remain the default choice for BMA in logistic regression, and their comparative performance is best assessed empirically.

All of the $g$-prior formulations evaluated in the linear setting are also available for logistic regression. This includes fixed-$g$ priors such as the unit information prior (with $g = n$), the $g = 4$ prior \citep{WuEtAl2016}, the so-called benchmark prior (with $g = \max\{n, p^2\}$), and $g = \sqrt{n}$, which has been shown to perform well in high-dimensional settings \citep{fernandez2001benchmark, young2014fast}. Other variants include the global and local empirical Bayes priors \cite{george2000calibration} and the hyper-$g$ prior, which adapt $g$ based on the data through either empirical estimation or full Bayesian modeling. These priors differ in how they balance model complexity and parsimony, and they imply different effective prior sample sizes. The unit information prior has a close connection to the Bayesian Information Criterion (BIC), which can be used to approximate marginal likelihoods and posterior model probabilities in logistic regression \citep{raftery1995bayesian}. Similarly, the Akaike Information Criterion (AIC) can be viewed as an approximation corresponding to $g = 1$ \citep{akaike1974new, burnham2002model}, although recent work has shown that AIC-based selection may perform poorly under separation \citep{kotani2025goodness}.

Several ways to extend $g$-priors to logistic regression models have been proposed \cite{SabanesBoveHeld2011,li_clyde_2018}. Here we focus on the proposal of Li and Clyde \cite{li_clyde_2018} to use a normal prior distribution with variance matrix proportional to the observed information matrix. Bayesian analysis of logistic regression models with $g$-priors is more complex than that of linear regression models due to the non-Gaussian likelihood. As a result, the implications of different choices of $g$ are less well understood in this setting. In the context of linear regression, Porwal and Raftery \citep{PorwalRaftery_2022} found that the priors with $g = \sqrt{n}$, local empirical Bayes, and the hyper-$g$ prior performed best across a diverse set of real-data scenarios. Whether these results carry over to logistic regression is an important question that we investigate in this article.

In logistic regression, prior specification plays an important role in settings where separation is present. Proper prior distributions can mitigate separation by regularizing extreme coefficient estimates and improving numerical stability. The use of proper but weakly informative priors is commonly recommended to address separation and improve robustness and interpretability, and heavy-tailed priors such as the Cauchy and the log-$F$ prior have been advocated for single-model inference \citep{gelman2008prior, greenland2016penalization}. The Cauchy prior is closely related to the Zellner–Siow and mixture $g$-prior constructions often used in Bayesian model averaging. However, recent work shows that Cauchy-like tail behavior may yield ill-behaved posterior moments or extreme estimates under separation, motivating interest in lighter-tailed alternatives \citep{GhoshLiMitra2018, Ghosh2019}. Priors with different tail behavior offer varying degrees of shrinkage, and their ability to stabilize inference under separation can differ substantially.

Several extensions of the $g$-prior framework have been implemented in BMA software for logistic regression, including the \texttt{BAS} package. The robust prior was introduced to reduce sensitivity to the choice of a fixed $g$ \citep{ Strawderman1971, bayarri2012criteria}. Both the robust and intrinsic priors truncate the support of $g$ in a manner that depends on model dimension and sample size. As a result, their effective support can differ substantially from that of local empirical Bayes, benchmark, or hyper-$g/n$ priors when the sample size is large relative to the number of predictors \citep{LeySteel2012, li_clyde_2018}. The Confluent Compound Hypergeometric (CCH) prior provides a unifying framework that encompasses these priors and their truncated variants as special cases \citep{li_clyde_2018}. Their behavior in logistic regression settings has received little systematic study.

In summary, most prior distributions used in Bayesian model averaging for logistic regression are extensions of priors originally developed for linear models. 
%While their use is widespread and supported in software, their theoretical properties under logistic likelihoods are not as well established.
Evaluating their performance under realistic conditions, including in the presence of separation, remains an important task.

In the frequentist setting, penalized likelihood approaches for logistic regression convert variable selection into an optimization problem. The objective typically involves the negative log-likelihood function with a penalty term $h_\lambda(\beta_1, \dots, \beta_p)$ on the coefficients $\boldsymbol{\beta}$. The penalized logistic regression estimate is
\begin{align*}
\hat{\beta} = \argmin_{\beta_1, \dots, \beta_p} &\left\lbrace -\sum_{i=1}^n \left[ y_i \cdot \eta_i - \log(1 + \exp(\eta_i)) \right]\right.\\ &+ \left. h_\lambda(\beta_1, \dots, \beta_p) \right\rbrace,
\end{align*}
where $\eta_i = \beta_0+ \sum_{j=1}^p \beta_j x_{i,j}$. As in linear regression, the estimates from these techniques can be interpreted as maximum a posteriori (MAP) estimates.

The LASSO \citep[Least Absolute Shrinkage and Selection Operator;][]{tibshirani1996regression}, where the penalty is $h_\lambda(\beta) = \lambda \sum_{j=1}^p |\beta_j|$, remains one of the most widely used methods. It encourages sparse solutions by shrinking some coefficients to exactly zero. Its popularity stems from its ability to perform variable selection while remaining computationally efficient in high-dimensional settings \citep{buhlmann2011statistics}. For logistic regression, efficient algorithms based on coordinate descent have been developed to optimize the penalized log-likelihood, thus scaling up LASSO to larger problems \citep{park2007l1, method_glmnet}.

However, the known limitations of the LASSO in linear regression also apply to logistic regression. The oracle property ensures that the true predictors will be included in the selected model asymptotically, but not that all selected predictors are truly relevant, even asymptotically. As a result, the LASSO can produce many false positives, even in large samples. It also tends to over-shrink large signals, which can lead to biased estimates \citep{kyung2010penalized}. The LASSO can also behave erratically in the presence of highly correlated predictors. It tends to select one variable from a group of collinear covariates and often discards the others, even if they carry important but weaker signals. This behavior can result in model instability and biased coefficient estimates, as relevant variables may be excluded due to their correlation with stronger predictors \citep{holmes2011discussion}.

Several extensions have been proposed to address these concerns. The ridge penalty \citep{hoerl1970ridge, lecessie1992ridge}, $h_\lambda(\beta) = \lambda \sum_{j=1}^p \beta_j^2$, does not induce sparsity, but it improves stability and prediction when predictors are highly correlated. The elastic net combines LASSO and ridge penalties to encourage group selection and stability across correlated features \citep{zou2005regularization}. Non-convex penalties such as SCAD \citep{fan2001variable} and MCP \citep{zhang2010nearly, breheny2011coordinate} reduce bias in large coefficients by applying weaker penalization at higher magnitudes. However, they are more difficult to optimize and are less commonly used in practice due to computational and tuning complexities. An alternative form of penalization is used in Firth's bias-reduced logistic regression, which applies a penalty derived from Jeffreys' invariant prior to reduce the first-order bias of the maximum likelihood estimator \citep{firth1993bias}. This method yields finite estimates even in the presence of complete separation and has been stated to be particularly useful in small-sample settings \citep{heinze2002solution}.

A common issue with penalized likelihood approaches is the lack of uncertainty quantification. In these approaches, variable selection is the outcome of a constrained optimization problem, and not a probabilistic statement of inclusion \citep{Park2008, fouskakis2015power, womack2014inference}. Consequently, the zeros induced may not correspond to the variables that would be excluded under a full probabilistic model selection framework \citep{hans2009bayesian}. These methods also provide no natural way to account for model uncertainty, which can lead to overconfident or unstable inference.

%% This should go to the discussion
Several methods have been proposed to address the lack of uncertainty quantification in penalized regression. One approach is the desparsified or debiased LASSO, which corrects the bias introduced by penalization and allows for asymptotically valid confidence intervals and hypothesis testing in high-dimensional settings \citep{van2014asymptotically, javanmard2014confidence}. A related alternative is the Induced Smoothed LASSO (IS-LASSO) \citep{method_islasso}, which replaces the non-differentiable $\ell_1$ penalty with a smooth approximation. This enables likelihood-based inference, such as Wald and score tests, while maintaining sparsity. Bayesian approaches, such as EM Variable Selection (EMVS) \citep{Rockova2014} and the spike-and-slab LASSO \citep{rockova2018spike}, combine ideas from model averaging and penalization to allow uncertainty quantification in principle. However, to our knowledge these ideas have not yet been implemented for logistic regression, and the ideas for uncertainty quantification are not yet fully supported in publicly available software.

While this wide range of methods has been proposed for variable selection and inference in logistic regression, it remains unclear which ones perform best across the diverse settings encountered in practice. Penalized likelihood approaches such as the LASSO are widely used due to their computational efficiency and accessibility through software like \texttt{glmnet} \citep{method_glmnet}, while Bayesian methods offer a principled framework for uncertainty quantification but are less commonly adopted. Beyond the challenge of selecting relevant predictors, empirical analyses often encounter issues such as separation, particularly in small samples or high-dimensional settings. These practical complications can compromise inference and highlight the need for systematic comparisons across methods.

To address these gaps, we conducted a preregistered study evaluating 28 established methods for statistical inference in logistic regression under model uncertainty. The full analysis pipeline, including the selection of methods, datasets, simulation design, and evaluation metrics, was specified in advance and made publicly available at \url{https://osf.io/swjcf} \citep{sekulovski2026comparing}. Our simulation design is based on real datasets spanning a variety of applied domains and involving practical situations in which the number of predictors may be much smaller than, comparable to, or substantially larger than the sample size. Full details are provided in the \emph{Materials and Methods} section and the \emph{Supporting Information}.

\section*{Results}

\begin{figure*}
\centering
\includegraphics[width=1\linewidth]{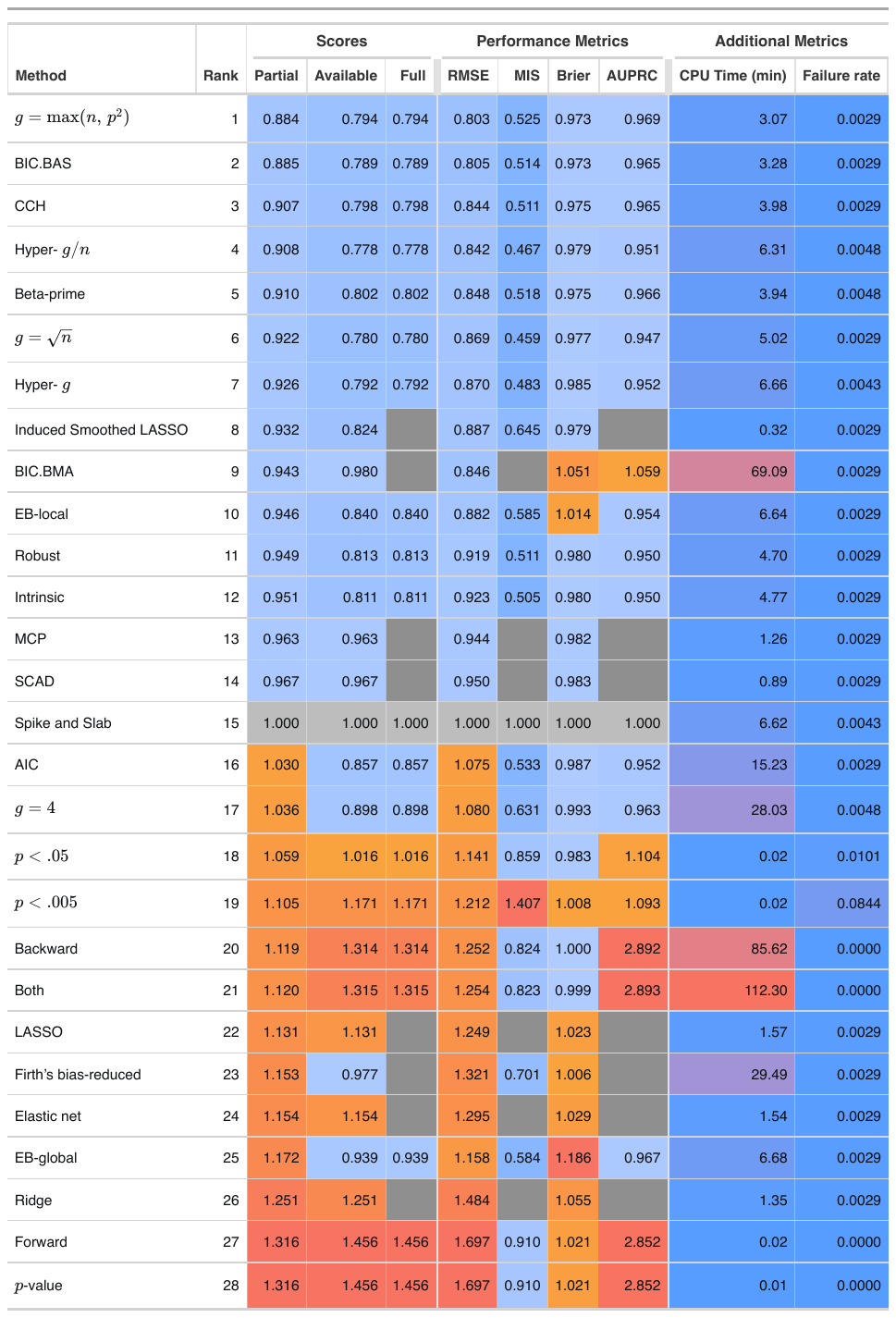}
\caption{Scores, performance, and additional metrics for 28 logistic regression methods with model uncertainty, averaged across 11 simulations on datasets without separation. Methods are ranked by the Partial Score, computed from RMSE (point estimation) and Brier score (prediction). All scores and performance metrics are standardized relative to the Spike–and–Slab method. Lower scores indicate better performance (blue), higher scores worse (orange/red). Additional metrics show average CPU time (minutes) and the proportion of failed models.}
\label{fig:res_nosep}
\end{figure*}

\begin{figure*}
\centering
\includegraphics[width=1\linewidth]{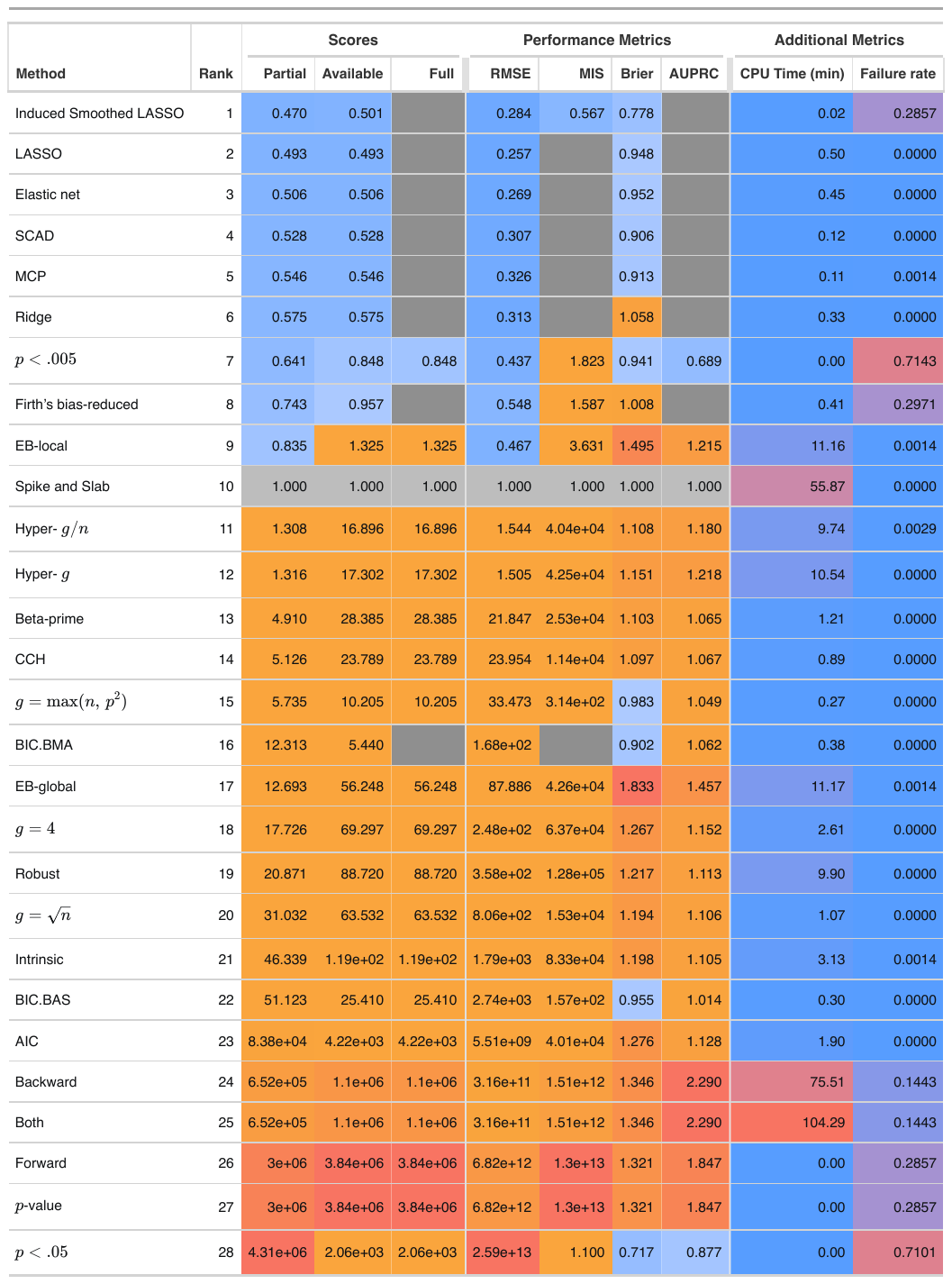}
\caption{Scores, performance, and additional metrics for 28 logistic regression methods with model uncertainty, averaged across 11 simulations on datasets with separation. Methods are ranked by the Partial Score, computed from RMSE (point estimation) and Brier score (prediction). All scores are standardized relative to the Spike–and–Slab method. Lower scores indicate better performance (blue), higher scores worse (orange/red). Additional metrics show average CPU time (minutes) and the proportion of failed models.
}
\label{fig:res_sep}
\end{figure*}

Fig.~\ref{fig:res_nosep} and Fig.~\ref{fig:res_sep} summarize the results. Fig.~\ref{fig:res_nosep} presents the results for datasets \emph{without} separation, whereas Fig.~\ref{fig:res_sep} shows the results for datasets \emph{with} separation. 
The tables report standardized performance metrics for point estimation (RMSE), interval estimation (IS), prediction (Brier), and overall model selection (AUPRC). All metrics are standardized such that lower values indicate better performance and are ranked relative to the Spike–and–Slab method. Ranking scores are then calculated from these standardized metrics.
The Full Score is defined as the average across all performance metrics, the Available Score as the average across the metrics available for each method, and the Partial Score as the average based on the RMSE and Brier score metrics, which are available for all methods. 
The methods are ranked according to the Partial Score. 
CPU time (in minutes) and failure rates are also reported, but are not used to compute the scores or determine the ranking. 
Simulation details are provided in the \emph{Materials and Methods} section and the \emph{Supporting Information}, and per-metric and per-dataset results are available in the \emph{Supporting Information}. In the \emph{Materials and Methods} section, we detail the discrepancies between the preregistration plan and the final analyses. The most notable deviation is that we did not preregister monitoring for separation, nor the decision to present results stratified by the presence or absence of separation. This was not anticipated at the preregistration stage, as we did not foresee the extent to which separation would influence the analysis.

Across datasets without separation (Fig.~\ref{fig:res_nosep}), Bayesian model averaging (BMA) methods from the \texttt{BAS} package generally perform best. Among these, the so-called benchmark prior with $g = \max(n, p^2)$  achieves the best scores overall, closely followed by the BIC.BAS, CCH, hyper--$g/n$, Beta-prime, and $g = \sqrt{n}$ priors. The Hyper-$g$, robust, intrinsic, and Spike and Slab priors also show strong and consistent performance across all metrics. The EB-local prior also performs competitively, though it slightly underperforms in prediction relative to the aforementioned priors. In contrast, EB-global performs poorly in both estimation and prediction, whereas the AIC and $g = 4$ methods show weaker performance in estimation; all three are ranked below the reference method. An additional examination of the posterior expected model size revealed that priors such as Hyper-$g$, EB.local, and $g=4$ tend to favor larger models, particularly in low-dimensional datasets, which partly explains their relatively weaker performance in estimation. In contrast, $g = \max(n, p^2)$ and BIC.BAS consistently yielded more parsimonious models across all datasets. All Bayesian methods exhibit a failure rate below 1\%. The BIC.BMA method implemented in the \texttt{BMA} package shows slightly poorer predictive accuracy and exhibits relatively slow computational performance on average.

Among the penalized regression approaches, the Induced Smoothed LASSO demonstrates the best overall performance (ranking at the 8th spot) and is also among the fastest methods. MCP and SCAD follow closely, performing well in both estimation and prediction. In contrast, the LASSO, Firth’s bias-reduced method, elastic net, and ridge regression tend to underperform relative to these methods. All these methods have failure rates below 1\%, and, apart from Firth’s method, are computationally efficient.

Finally, the classical stepwise and $p$-value–based selection methods perform notably worse than the Bayesian and penalized regression approaches overall. The $p < .05$ and $p < .005$ thresholds yield poor results, with the latter also having a 8\% failure rate. The forward and backward stepwise approaches perform similarly poorly and are the slowest among all 28 methods considered.

The results for datasets with and without separation differ greatly (cf. Fig.~\ref{fig:res_sep}). Across datasets with separation, penalized likelihood methods, such as the Induced Smoothed LASSO, LASSO, elastic net, SCAD, MCP, and ridge regression consistently demonstrate strong performance across all available metrics. However, the Induced Smoothed LASSO (ranked first) had a failure rate of 28.5\%, and Firth's bias-reduced method (ranked eighth) had a failure rate of 29.7\%. Therefore, their rankings should be interpreted with caution, as the metrics of the other methods take into account the performance on datasets on which these two methods failed.

Among the Bayesian approaches, the EB-local and Spike–and–Slab methods stand out due to their robust estimation and interval calibration. The EB-local method ranked ninth and the Spike-and-Slab method ranked tenth. However, the Spike-and-Slab method tends to be relatively slow. Several $g$-prior methods, as well as the AIC method, demonstrate significant performance degradation, particularly in point and interval estimation. However, some methods, such as $g = \max(n, p^2)$, BIC.BAS, and BIC.BMA, maintain relatively strong performance in prediction and overall model selection, despite underperforming in estimation. All Bayesian methods had failure rates below 1\%.

The $p$-value-based and stepwise selection methods now firmly rank among the worst performers. The apparent exception is the $p < .005$ method, but its 71\% failure rate renders this ranking unreliable. All of the $p$-value and stepwise selection methods had high failure rates. The Both and Forward stepwise methods are also among the slowest of the evaluated methods.

\section*{Discussion}

% Our study
Previous comparative studies of variable selection methods in logistic regression have been limited in scope, considering only a small subset of available approaches or using simulation designs not closely tied to empirical data. In contrast, our study provides a comprehensive, preregistered comparison of 28 publicly available methods for variable selection and inference under model uncertainty in logistic regression. We based 11 simulation studies directly on publicly available datasets to ensure that our design reflects the realistic data structures encountered in applied research.

Two features distinguish our study from earlier comparisons. First, preregistration of the study design, including the selection of datasets, methods, and metrics and the specification of data-generating models, reduced researcher degrees of freedom and increased transparency. When we observed separation in several simulation conditions, we adapted our analysis plan accordingly. We also made transparent how these adaptations were implemented, including the procedure used to identify which datasets exhibited separation and which did not. This allowed us to report results separately for datasets with and without separation, a distinction that has not been considered in previous comparisons. Second, our simulation design employed parametric bootstrapping to approximate the data-generating processes underlying real datasets, thereby increasing the empirical grounding of our findings relative to studies that rely on fully synthetic data-generating models. Together, these features make our comparison both broader in scope and closer to practical data-analysis settings than earlier work.

% Relate our results to earlier work
Our study is a conceptual replication of the work of Porwal and Raftery \citep{PorwalRaftery_2022}, who conducted perhaps the most extensive comparison to date of methods for variable selection under model uncertainty, focusing on linear regression. Their simulations, which were also based directly on real datasets, showed that adaptive Bayesian model averaging (BMA) methods using $g$-priors performed best across estimation, inference, and prediction tasks. By adaptive BMA methods, we mean BMA methods in which the parameter priors and/or the model space prior are themselves estimated from the data. More specifically here, when $g$-priors are used, it refers to methods in which $g$ is estimated from or otherwise adapts to the data. This should not be confused with the adaptive MCMC algorithms sometimes used in Bayesian statistics.

The $g=\sqrt{n}$ and empirical Bayes local (EB-local) priors performed particularly well in that study \citep{PorwalRaftery_2022}. Our design closely follows theirs but extends it to logistic regression, where separation and nonlinearity present additional challenges. The strong performance of $g$-prior-based and EB-local BMA methods in our study suggests that the advantages of adaptive priors extend beyond linear models. However, since separation does not arise in linear models, our results also reveal an additional strength of the EB-local method: it remains robust across models and data conditions, including those affected by separation.

Many comparisons of variable selection methods have been made for linear regression, but few have been made specifically for logistic regression, where issues such as separation and nonlinearity introduce additional challenges. One of the most extensive earlier comparisons was carried out by Li and Clyde \citep{li_clyde_2018}, who evaluated a broad class of $g$-priors using simulated datasets with correlated and uncorrelated predictors and varying levels of sparsity in the generating model. The priors they examined included hyper-$g$, hyper-$g/n$, beta-prime, CCH, robust, intrinsic, benchmark, and EB-local priors. All of these priors were also part of our analysis. Li and Clyde found that no single prior performed best across all criteria; some variants performed better for model selection, while others performed better for prediction. However, their simulations were based on synthetic designs and did not include separation or comparisons with non-$g$-prior methods. Our study complements theirs by evaluating these priors alongside additional Bayesian and penalized likelihood methods in empirically grounded simulations. Consistent with their results, we found that adaptive and mixture-based $g$-priors perform well under no separation. However, we also observed a marked decline in their performance under separation, a scenario Li and Clyde did not consider. In this setting, the EB-local prior maintained strong performance across metrics, further supporting its value as a versatile default for logistic regression.

Two smaller studies also compared methods for logistic regression and provide useful points of reference. Wang, Zhang, and Bakhai \citep{wang2004comparison} compared BMA with classical stepwise procedures in two small simulation settings ($n = 500$, $p = 25$). In both settings, one with no true predictors and one with two predictors, BMA identified the correct model consistently, while stepwise procedures tended to select spurious predictors. Our results are fully consistent with theirs. BMA methods outperformed stepwise selection across all conditions. Wu, Ferreira, and Gompper \citep{WuEtAl2016} compared the fixed $g = 4$ prior, the hyper-$g/n$ prior, and BIC in simulations with eight predictors and varying sample sizes. Overall, the three methods performed similarly. The $g$-prior variants performed slightly better when predictors were present in the data-generating model, while BIC performed better for the empty data-generating model. Our results for the inference metric (AUPRC) are similar to theirs, but we also found that the fixed $g = 4$ prior performed worse than the hyper-$g/n$ and BIC.BAS approaches in terms of estimation metrics (RMSE and MIS) in the non-separation scenario. This suggests that estimation accuracy is an important dimension of performance for $g$-priors, particularly under separation.

Indeed, the sharp decline in the overall performance of $g$-prior-based methods for datasets showing separation is primarily driven by reduced point and interval estimation accuracy, as indicated by higher root mean squared errors and mean interval scores. Nevertheless, inference and predictive performance remain strong, with relatively stable areas under the precision–recall curve and Brier scores. This pattern aligns with the theoretical results of Li and Clyde \citep{li_clyde_2018}, who showed that posterior distributions under $g$-priors may not be well-defined when maximum likelihood estimates do not exist (as in the case of separation). However, model-averaged inference and predictive quantities remain valid.

Several other Bayesian variable selection methods for logistic regression have recently been proposed. However, we did not include them in our comparison because we could not find any publicly available software implementations. These methods include the power-expected-posterior prior of Fouskakis, Ntzoufras, and Perrakis \citep{Fouskakis2018}; the nonlocal prior approach of Shi, Lim, and Maiti \citep{Shi2019}; the variational Bayes method of Zhang, Xu, and Zhang \citep{Zhang2019}; and the environmental DNA–motivated model of Griffin et al. \citep{Griffin2020}. For similar reasons, we did not include the test-based Bayes factor methods for probit regression proposed by Hu and Johnson \citep{HuJohnson2009} and for logistic regression by Held, Saban\`{e}s Bov\`{e}, and Gravestock \citep{HeldSabanesBoveGravestock2015}. Earlier work by Liao and Chin \citep{LiaoChin2007} proposed a parametric bootstrap method tailored to high-dimensional ($p \gg n$) settings. This method was implemented in the \texttt{GeneLogit} package, which is no longer available. Similarly, methods using the Laplace Power Exponential Prior (LPEP) were shown in simulations to be competitive \citep{Porwal_2024}, although an R package on CRAN is not available at this time. Once accessible implementations exist, evaluating these and other emerging methods would further strengthen comparative assessments of Bayesian variable selection for logistic regression.

Most of the Bayesian methods we have considered involve model averaging, while most of the frequentist methods focus on model selection. This reflects the historical development of the area. However, in the past decade there has been increasing attention paid to frequentist model averaging (FMA) \cite{HjortClaeskens2003,ClaeskensHjort2008,Steel2020}. FMA itself involves many choices and hence has many variants, and the associated publicly available software is less developed than for the methods we have focused on here. However, it would be of interest to extend the present study to include FMA methods.

Our analysis also focused narrowly on logistic regression methods. We did not compare logistic regression with other machine learning algorithms because doing so would have expanded the scope of the study considerably. Song et al. \citep{Song2021}, who reviewed 24 studies comparing logistic regression with various machine learning methods for predicting acute kidney injury, found broadly similar performance between the two approaches, though there was greater variability among machine learning methods. Additionally, Sur and Cand\`{e}s \citep{SurCandes2019} showed that standard large-sample theory for significance testing in logistic regression breaks down when the number of variables is large, even if it remains smaller than the sample size. They proposed an alternative asymptotic framework. It would be valuable to examine how the methods evaluated here behave under those high-dimensional conditions.

Finally, our comparison addressed model uncertainty arising from uncertainty about which variables to include, but not other forms of model uncertainty such as  alternative link functions like the complementary log-log function. Extending the present framework to these settings would provide a more complete picture of model uncertainty in generalized linear models. Likewise, future work could explore the impact of different model space priors. In this study, we used a truncated beta-binomial (1,1) prior that assigns equal probability across model sizes up to $n - 2$. This choice of model space prior may partly explain the observed tendency toward larger posterior expected model sizes observed for some BAS methods in low-dimensional datasets. Therefore, alternative priors that more strongly penalize model complexity or encode prior information about sparsity could lead to different conclusions. Finally, it should be noted that the performance differences between BIC.BMA and BIC.BAS partly reflect differences in their model space priors and search algorithms.

Note that, while the datasets in our study are based on real data, the actual observations are simulated
from a model in the set of all possible models being considered. Thus the study is in the context of the M-closed framework \cite{BernardoSmith1994}. One implication is that the generating model is linear in a set of considered predictors. This is more general than one might think, however. For example, nonlinear effects can be included in the framework by including regressors that are nonlinear transformations of regressors or interactions.  

In summary, our study provides a comprehensive, preregistered, empirical comparison of 28 established methods for variable selection and inference under model uncertainty in logistic regression. The EB-local prior was the most robust method across separation and non-separation conditions, offering a balance of accuracy, stability, and computational efficiency. Penalized methods, such as LASSO, SCAD and MCP, are strong alternatives when separation is likely and quantifying model uncertainty is deemed unnecessary. Together, these findings and the preregistered, data-based design of the study demonstrate the feasibility of transparent, reproducible simulation studies in statistical methodology.

\section*{Materials and Methods}
We conducted 11 simulations based closely on 11 empirical datasets. In the following subsections, we describe the key decisions and methodological choices underlying our approach. Additional details on each empirical dataset used in the simulations, as well as extended results, are provided in the Supporting Information. The simulation study was preregistered at \url{https://osf.io/swjcf}. In each of the following subsections, we detail any deviations from the submitted preregistration plan.

\subsection*{Statistical Methods for Variable Selection}
Here we outline the variable selection methods we compared and explain some of the choices we made.
We began by exploring which R packages were available for logistic regression.
To this end, we consulted the CRAN Task Views for Bayesian Inference and Machine Learning, and additionally searched the source code of all CRAN Task Views for the term “Logistic regression” (\href{https://github.com/search?q=org%3Acran-task-views+logistic+regression&type=code}{GitHub-link}).
We considered a package potentially suitable for inclusion if it was capable of performing logistic regression and offered some form of variable selection, regardless of the specific type.
Methods that did not rely on a logistic link function (e.g., probit regression) were excluded.

\begin{table*}[!ht]
    \caption{Overview of the methods compared in the simulation study}
    \label{tb:methods_overview}
    \centering
    \footnotesize
    \setlength{\tabcolsep}{4pt}
    \resizebox{\textwidth}{!}{%
    \begin{tabular}{llccl}
    \toprule
    Method & Type & References & Implementation & Function \\
    \midrule
    % BAS
    Hyper$-g$               & Bayesian & \cite{liang2008mixtures, SabanesBoveHeld2011} & BAS-V1.7.5 & \texttt{bas.glm(..., betaprior = hyper.g())}  \\
    Hyper$-g/n$             & Bayesian & \cite{liang2008mixtures, SabanesBoveHeld2011} & BAS-V1.7.5 & \texttt{bas.glm(..., betaprior = hyper.g.n())}  \\
    $g =\sqrt n$            & Bayesian & \cite{fernandez2001benchmark}                 & BAS-V1.7.5 & \texttt{bas.glm(..., betaprior = g.prior(sqrt(n)))}  \\
    $g = \max(n,\, p^2)$    & Bayesian & \cite{fernandez2001benchmark}                 & BAS-V1.7.5 & \texttt{bas.glm(..., betaprior = g.prior(max(n, p\^2)))}  \\
    $g=4$                   & Bayesian & \cite{WuEtAl2016}                             & BAS-V1.7.5 & \texttt{bas.glm(..., betaprior = g.prior(4))}  \\
    CCH                     & Bayesian & \cite{li_clyde_2018}                          & BAS-V1.7.5 & \texttt{bas.glm(..., betaprior = CCH(1, n, 0))}  \\
    Robust                  & Bayesian & \cite{bayarri2012criteria}                    & BAS-V1.7.5 & \texttt{bas.glm(..., betaprior = robust())}  \\
    Intrinsic               & Bayesian & \cite{Berger1996}                             & BAS-V1.7.5 & \texttt{bas.glm(..., betaprior = intrinsic())}  \\
    Beta-prime              & Bayesian & \cite{Maruyama2011}                           & BAS-V1.7.5 & \texttt{bas.glm(..., betaprior = beta.prime())}  \\
    EB-local                & Bayesian & \cite{clyde2000flexible, george2000calibration} & BAS-V1.7.5 & \texttt{bas.glm(..., betaprior = EB.local())}  \\
    EB-global               & Bayesian & \cite{clyde2000flexible, george2000calibration} & BAS-V1.7.5 & \texttt{bas.glm(..., betaprior = EB.global())}  \\
    AIC                     & Bayesian & \cite{akaike1974new, burnham2002model, li_clyde_2018} & BAS-V1.7.5 & \texttt{bas.glm(..., betaprior = aic.prior())}  \\
    BIC.BAS                 & Bayesian & \cite{raftery1995bayesian, li_clyde_2018, Raftery_1996_approx} & BAS-V1.7.5 & \texttt{bas.glm(..., betaprior = bic.prior())}  \\
    % glmnet
    Ridge                   & Frequentist & \cite{hoerl1970ridge, lecessie1992ridge}       & glmnet-V4.1.9   & \texttt{cv.glmnet(..., alpha = 0)}   \\
    Elastic net             & Frequentist & \cite{zou2005regularization}                  & glmnet-V4.1.9   & \texttt{cv.glmnet(..., alpha = 1/2)} \\
    LASSO                   & Frequentist & \cite{tibshirani1996regression}               & glmnet-V4.1.9   & \texttt{cv.glmnet(..., alpha = 1)}   \\
    % stats
    $p-$value               & Frequentist & \cite{HosmerLemeshow2013}                     & stats-V4.5.0    & \texttt{glm(...)}   \\
    $p<.05$                 & Frequentist & \cite{HosmerLemeshow2013}                     & stats-V4.5.0    & \texttt{glm(...)}   \\
    $p<.005$                & Frequentist & \cite{HosmerLemeshow2013}                     & stats-V4.5.0    & \texttt{glm(...)}   \\
    Forward                 & Frequentist & \cite{HosmerLemeshow2013}                     & stats-V4.5.0    & \texttt{step(glm(...), direction = "forward")}   \\
    Backward                & Frequentist & \cite{HosmerLemeshow2013}                     & stats-V4.5.0    & \texttt{step(glm(...), direction = "backward")}  \\
    Both                    & Frequentist & \cite{HosmerLemeshow2013}                     & stats-V4.5.0    & \texttt{step(glm(...), direction = "both")}      \\
    % BoomSpikeSlab
    Spike and Slab          & Bayesian    & \cite{Tuchler01032008, method_BoomSpikeSlab}  & BoomSpikeSlab-V1.2.6 & \texttt{logit.spike(..., prior = LogitZellnerPrior(...))} \\
    % ncvreg
    MCP                     & Frequentist & \cite{zhang2010nearly, breheny2011coordinate} & ncvreg-V3.15.0  & \texttt{cv.ncvreg(..., penalty = "MCP")}  \\
    SCAD                    & Frequentist & \cite{fan2001variable}                        & ncvreg-V3.15.0  & \texttt{cv.ncvreg(..., penalty = "SCAD")} \\
    % islasso
    Induced Smoothed LASSO  & Frequentist & \cite{method_islasso}                         & islasso-V1.5.2  & \texttt{islasso(...)} \\
    % logistf
    Firth's Bias-Reduced    & Frequentist & \cite{firth1993bias, heinze2002solution}       & logistf-V1.26.1 & \texttt{logistf(...)} \\
    % BMA
    BIC.BMA                 & Bayesian    & \cite{raftery1995bayesian}                    & BMA-V3.18.20    & \texttt{bic.glm(...)}  \\
    \bottomrule
    \end{tabular}}
\end{table*}
Table~\ref{tb:methods_overview} provides an overview of the included methods.  By methods, we also refer to different implementation options within R packages; for instance, the \texttt{BAS} package offers several parameter priors that have been shown to yield varying performance \citep{li_clyde_2018, PorwalRaftery_2022}, and we therefore treat each of these priors as a distinct method. For each method we provide the generic function call used in our simulations; for Bayesian methods this also indicates the parameter prior distribution used, with the exception of the \texttt{BMA} package, where default prior settings are used. We did not vary the model space prior for any of the Bayesian methods. For all the methods in the \texttt{BAS} package we fixed the model space prior to a truncated beta-binomial($\alpha = 1,\, \beta = 1$) so that models with $n - 2$ or more predictors received a prior probability of 0; for datasets with $n > p$ this reduces to an untruncated beta-binomial prior that is uniform over model size. For the Spike and Slab method we used the default options in the \texttt{LogitZellnerPrior()} function with \texttt{prior.success.probability~=~0.5} and \texttt{expected.model.size~=~1}, 
which assigns each variable a prior inclusion probability of $\min(1, 1/p)$. For BIC.BMA we used the default prior inclusion probability of $0.5$ for each variable.
In the preregistration, we indicated that we would conduct initial simulation studies to determine which model-space priors to use. However, due to computational considerations, we ultimately decided to retain the default choices described above.

We used the default MCMC and optimization algorithms for all the R packages.
For the sake of computational efficiency, we limited the number of posterior samples for \texttt{BAS} and \texttt{BoomSpikeSlab} to 10,000, matching the number of samples used in earlier large scale simulation studies \cite[e.g.,][]{PorwalRaftery_2022, porwal2022_modelpriors}.

\subsection*{Datasets}
For the simulation study, we selected 11 publicly available empirical datasets from various fields, including medicine, chemistry, genetics, the social sciences, and astronomy. These datasets represent a range of sample sizes ($n$) and numbers of covariates ($p$), including one dataset in a high-dimensional setting ($p > n$). We describe the full screening process in the Supporting Information.

\begin{figure}
\centering
\includegraphics[width=.7\linewidth]{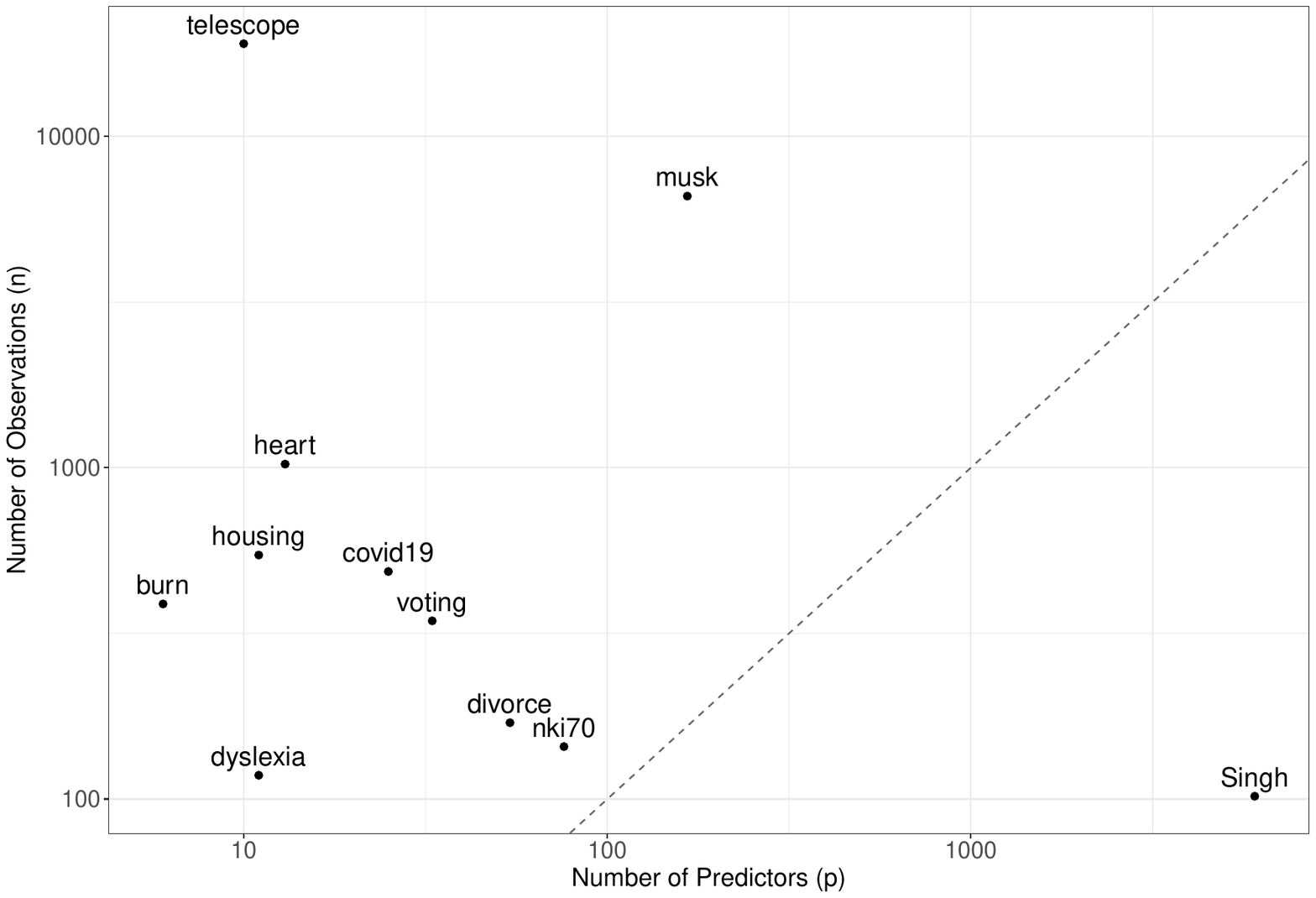}
\caption{Relationship between the number of predictors (p) and the number of observations (n) across the empirical data sets used in the simulations. Both axes are plotted on a log 10 scale to aid interpretability, but tick marks reflect the original (untransformed) values. The dashed line indicates the identity line (n = p)}
\label{fig:n_p}
\end{figure}

The final 11 datasets are listed in Table~\ref{tab:dataset_overview}. Figure~\ref{fig:n_p} displays the distribution of sample sizes and numbers of predictors across these datasets. As shown, most datasets fall within the low-predictor, moderate-sample-size range. In addition, the \emph{telescope} dataset contains a very large number of observations, the \emph{musk} dataset features both a large number of observations and a moderately large number of predictors, and the \emph{Singh} dataset represents a high-dimensional case with many predictors and a relatively small sample size. We argue that this composition reasonably reflects the distribution of datasets encountered in applied research and matches the preregistered target of conducting approximately 10–15 simulation scenarios. 

\begin{table}
\caption{Overview of the final selected empirical datasets, including sample size ($N$), number of predictors ($P$), and the corresponding source.}
\label{tab:dataset_overview}
\centering
\begin{tabular}{lccl}
\toprule
\textbf{Name} & \textbf{N} & \textbf{P} & \textbf{Source} \\ \midrule
burn & 388 & 6 & \cite{HosmerLemeshow2013,R-aplore3} \\ 
telescope & 19,020 & 10 & \cite{mitgandhi2023logistic} \\
dyslexia & 118 & 11 & \cite{Dbska2021,osfDbska2021} \\
housing & 545 & 11 & \cite{yasserh2023housing} \\
heart & 1,025 & 13 & \cite{johnsmith88_heart_disease} \\
covid19 & 468 & 25 & \cite{osf-covid19} \\
voting & 345 & 34 & \cite{Brandt-2021} \\
divorce & 170 & 54 & \cite{UCIdivorce} \\
nki70 & 144 & 70 & \cite{nki70-R-penalized} \\
musk & 6,598 & 166 & \cite{chapman1994musk} \\
Singh & 102 & 6,033 & \cite{singh2002} \\
\bottomrule
\end{tabular}
\end{table}

\subsection*{Data Generating Process}

For each empirical dataset, we derived a data-generating model following the procedure described in the Supporting Information, and simulated 100 binary outcome vectors while holding the predictor matrix fixed. Each simulated dataset was then analyzed using all 28 candidate methods (Table~\ref{tb:methods_overview}) with 5-fold cross-validation.

We monitored whether separation occurred in any dataset at any fold of the cross-validation using the R package \texttt{detectseparation} \citep{detectseparation_package}. Out of the 1,100 simulated datasets, 463 (42\%) exhibited separation. For the \emph{covid19} dataset, 34 out of 100 simulated datasets showed separation; for \emph{heart}, 28 out of 100; and for \emph{housing}, 1 out of 100. All simulated datasets showed separation for the \emph{divorce}, \emph{nki70}, \emph{voting}, and \emph{Singh} datasets, whereas none exhibited separation for the \emph{burn}, \emph{dyslexia}, \emph{musk}, and \emph{telescope} datasets.

We note two divergences from the preregistration plan. First, the preregistration did not mention the use of 5-fold cross-validation. Second, and more importantly, we did not preregister the decision to report results separately based on whether separation occurred. This latter choice was made because we had not anticipated this issue prior to running the first of the 11 simulations.

A complete absence of results (i.e., total failure) for an entire simulation based on a specific dataset for a method occurred only in cases where the data exhibited separation. For the \textit{Singh} dataset, all replications failed for BIC.BMA, the Induced Smoothed Lasso, Firth’s Bias-Reduced method, and all stepwise and $p$-value--based methods (nine methods in total). For the \textit{Voting} dataset, all replications failed for BIC.BMA, the Induced Smoothed Lasso, Firth’s Bias-Reduced method, forward selection, and all $p$-value--based methods (seven methods in total). For the \textit{divorce} and \textit{nki70} datasets, the $p < 0.05$ and $p < 0.005$ methods failed completely (two methods each). 
Finally, for the \emph{covid19} dataset, the $p < 0.05$ method failed. Overall, these failures accounted for 7.6\% of the entire simulation and 19.2\% of all instances involving separation.

The performance measures used to compare the methods are described below. Each measure was averaged across the 100 repetitions for each simulation based on the 11 empirical datasets. We also report two additional metrics that, while not used for ranking, are important for interpreting the results.

In Figs.~\ref{fig:res_nosep} and \ref{fig:res_sep}, all performance metrics were first averaged over repetitions for each dataset–method pair, then standardized as log-ratios relative to a fixed reference method (the Spike–and–Slab). These standardized values were exponentiated for display, such that a score of 1 denotes parity with the reference method, values below 1 indicate better performance, and values above 1 indicate worse performance. The overall \emph{Full Score} corresponds to the geometric mean across all performance metrics. The \emph{Available Score} is computed on the subset of metrics provided by each method, and the \emph{Partial Score} is restricted to the core metrics shared across all methods (RMSE and prediction Brier score). Methods are ranked according to the \emph{Partial Score}. The additional metrics were not standardized.

\subsubsection*{Performance Metrics}

\begin{itemize}
	\item \textbf{Root Mean Squared Error (RMSE)}: The Root Mean Square Error (RMSE) measures, on average, how far the estimated coefficients are from the true coefficients (associated to the DGM). A lower RMSE indicates that the estimated coefficients are more accurate, where
		\[
        \text{RMSE} = \sqrt{\frac{1}{p}\sum_{j=1}^{p}{\left(\hat{\beta}_{j}-\beta^{\text{DGM}}_j\right)^{2}}} .
        \]
	\item \textbf{Mean Interval Score (MIS)}: The interval score \cite{Gneiting2007} is a measure to assess the quality of confidence/credible intervals for the coefficients. It is calculated as the interval width penalized by the missed coverage when the true value lies outside the interval. The smaller interval score the better.
    \[
    IS =
        \left\{ 
            \begin{array}{ll}
                \displaystyle u - l + \frac{2}{\alpha}(l - \beta^{\text{DGM}}) & \text{if } \beta^{\text{DGM}} < l \\
                \displaystyle u - l & \text{if } l \leq \beta^{\text{DGM}} \leq u \\
                \displaystyle u - l + \frac{2}{\alpha}(\beta^{\text{DGM}} - u) & \text{if } \beta^{\text{DGM}} > u , \\
            \end{array} 
        \right.
    \]
    where $l$ and $u$ are respectively the lower and upper bound of the confidence/credible interval, and $2/\alpha$ is a penalization coefficient for the missed coverage, with $\alpha = 0.05$. We compute the mean interval score (MIS) for all coefficients in a data set.
	\item \textbf{Area Under the Precision-Recall Curve (AUPRC)}: This metric evaluates the overall quality of model selection by assessing how effectively it identifies truly relevant variables among those that have been selected, without relying on a threshold for inclusion probability. 
    %It is particularly useful in sparse situations, where there are only a few relevant variables. 
    A higher Area Under the Precision-Recall Curve (AUPRC) indicates better identification of relevant variables while minimizing the number of false positives.
	%\item \textbf{Prediction $R^2$}: A measure to quantify the proportion of variation in binary outcomes explained by the predicted probabilities of the model, using a squared error metric that is similar to the $R^2$ in linear regression. Higher values indicate better predictive performance.
    \item \textbf{Brier score}: The Brier score \citep{brier1950verification} is a measure used to quantify the accuracy of binary probabilistic predictions on held-out test data. It is defined as the mean squared difference between predicted probabilities and the actual binary outcomes. Lower Brier scores indicate better calibrated and more accurate predictions:

    \begin{equation}
\text{BS} = \frac{1}{N} \sum_{i=1}^{N} (p_i - o_i)^2
\end{equation}
\end{itemize}

\subsubsection*{Additional Metrics}

\begin{itemize}
	\item \textbf{CPU time}: The time in seconds taken to run an analysis on one dataset using a specific method, (including all cross-validation folds). All methods were run on a single CPU core (with OpenMP and OpenBLAS threading disabled), so no parallelization was used. Methods relying on compiled code internally, still benefit from optimized compiled routines but not from parallelism. As already mentioned, MCMC-based methods (\texttt{BAS}, \texttt{BoomSpikeSlab}) were given a fixed budget of 10,000 iterations; no formal convergence diagnostics were applied, so runtimes for these methods reflect a fixed computational budget rather than a convergence-equivalent comparison with the optimization-based methods. The values presented in Figures \ref{fig:res_nosep} and \ref{fig:res_sep} are in minutes. This metric is included as an indicative measure of computational demand; given differences in implementation, algorithmic type, and computational budget, it should not be interpreted as a definitive basis for speed comparison across methods.
    
    \item \textbf{Failure rate}: The proportion of failed models, defined as the proportion of analyses that did not produce any usable output and exited with an error (i.e., complete failure).
\end{itemize}

We note several divergences from the preregistered analysis with respect to the performance metrics. First, although we preregistered the use of prediction $R^2$ (defined as the proportion of variation in binary outcomes explained by the model’s predicted probabilities, using a squared-error metric analogous to linear regression), we ultimately used the Brier score instead, as it is a more natural and widely accepted measure for binary outcomes.

Second, while the preregistration stated that both the Area Under the ROC Curve (AUC) and the Area Under the Precision–Recall Curve (AUPRC) would be used, we report only the AUPRC. This change was made due to redundancy between the two measures and because the AUPRC is more informative for evaluating model recovery in unbalanced settings.

Third, we chose to report the proportion of models that failed only after completing the simulations, as this metric emerged as relevant during the analysis process.

Finally, the exact approach for ranking methods and transforming performance metrics was not preregistered. However, we did preregister that performance metrics would be standardized and ranked relative to a reference method, which we adhered to.

\subsection*{Code \& Data Availability}

The code, data, preregistration document and supporting information are available at: \url{https://osf.io/swjcf}

\subsection*{Acknowledgments}

NS, DvdB, GA, and MM were supported by the European Union {(ERC, BAYESIAN P-NETS, \#101040876).

\bibliographystyle{unsrtnat}  % keeps references in order of first citation
\bibliography{pnas-sample}    % do NOT include .bib here

@misc{sekulovski2026comparing,
  author       = {Sekulovski, Nikola},
  title        = {Comparing Variable Selection and Model Averaging Methods for Logistic Regression},
  year         = {2026},
  month        = {October},
  day          = {22},
  howpublished = {\url{https://osf.io/swjcf}},
  note         = {Retrieved from OSF}
}

@article{Maruyama2011,
  title = {Fully {B}ayes factors with a generalized g-prior},
  volume = {39},
  ISSN = {0090-5364},
  url = {http://dx.doi.org/10.1214/11-aos917},
  DOI = {10.1214/11-aos917},
  number = {5},
  journal = {Annals of Statistics},
  publisher = {Institute of Mathematical Statistics},
  author = {Maruyama,  Yuzo and George,  Edward I.},
  year = {2011},
  month = oct,
  pages={2740--2765}
}

@article{Berger1996,
  title = {The Intrinsic {B}ayes Factor for Model Selection and Prediction},
  volume = {91},
  ISSN = {1537-274X},
  url = {http://dx.doi.org/10.1080/01621459.1996.10476668},
  DOI = {10.1080/01621459.1996.10476668},
  number = {433},
  journal = {Journal of the American Statistical Association},
  publisher = {Informa UK Limited},
  author = {Berger,  James O. and Pericchi,  Luis R.},
  year = {1996},
  pages = {109–122}
}

@Manual{detectseparation_package,
    title = {detectseparation: Detect and Check for Separation and Infinite Maximum Likelihood
Estimates},
    author = {Ioannis Kosmidis and Dirk Schumacher and Florian Schwendinger},
    year = {2022},
    note = {R package version 0.3},
    url = {https://CRAN.R-project.org/package=detectseparation},
    doi = {10.32614/CRAN.package.detectseparation},
  }

@article{li_clyde_2018,
  title={Mixtures of g-priors in generalized linear models},
  author={Li, Yingbo and Clyde, Merlise A},
  journal={Journal of the American Statistical Association},
  volume={113},
  number={524},
  pages={1828--1845},
  year={2018},
  doi={10.1080/01621459.2018.1469992}
}

@article{porwal2022_modelpriors,
  title={Effect of model space priors on statistical inference with model uncertainty},
  author={Porwal, Anupreet and Raftery, Adrian E},
  journal={New England Journal of Statistics in Data Science},
  volume={1},
  number={2},
  pages={149},
  doi={10.51387/22-NEJSDS14},
  year={2022}
}

@article{Song2021,
  title = {Comparison of machine learning and logistic regression models in predicting acute kidney injury: A systematic review and meta-analysis},
  volume = {151},
  ISSN = {1386-5056},
  url = {http://dx.doi.org/10.1016/j.ijmedinf.2021.104484},
  DOI = {10.1016/j.ijmedinf.2021.104484},
  journal = {International Journal of Medical Informatics},
  publisher = {Elsevier BV},
  author = {Song,  Xuan and Liu,  Xinyan and Liu,  Fei and Wang,  Chunting},
  year = {2021},
  month = jul,
  pages = {104484}
}

@article{Stoltzfus2011,
  author  = {Stoltzfus, John C.},
  title   = {Logistic Regression: A Brief Primer},
  journal = {Academic Emergency Medicine},
  year    = {2011},
  volume  = {18},
  number  = {10},
  pages   = {1099--1104},
  doi     = {10.1111/j.1553-2712.2011.01185.x},
}

@article{LiaoChin2007,
  author  = {Liao, Jiann-Gwo and Chin, Kelvin V.},
  title   = {Logistic regression for disease classification using microarray data: model selection in a large p and small n case},
  journal = {Bioinformatics},
  year    = {2007},
  volume  = {23},
  number  = {15},
  pages   = {1945--1951},
  doi     = {10.1093/bioinformatics/btm287},
}

@article{Bursac2008,
  author  = {Bursac, Zoran and Gauss, Charles H. and Williams, Deborah K. and Hosmer, David W.},
  title   = {Purposeful Selection of Variables in Logistic Regression},
  journal = {Source Code for Biology and Medicine},
  year    = {2008},
  volume  = {3},
  pages   = {17},
  doi     = {10.1186/1751-0473-3-17},
}

@article{SurCandes2019,
  author       = {Pragya Sur and Emmanuel J. Cand{\`e}s},
  title        = {A modern maximum-likelihood theory for high-dimensional logistic regression},
  journal      = {Proceedings of the National Academy of Sciences of the United States of America},
  year         = {2019},
  volume       = {116},
  number       = {29},
  pages        = {14516--14525},
  doi          = {10.1073/pnas.1810420116},
}

@article{singh2002,
  title={Gene expression correlates of clinical prostate cancer behavior},
  author={Singh, Dinesh and Febbo, Phillip G and Ross, Kenneth and Jackson, Donald G and Manola, Judith and Ladd, Christine and Tamayo, Pablo and Renshaw, Andrew A and D'Amico, Anthony V and Richie, Jerome P and others},
  journal={Cancer cell},
  volume={1},
  number={2},
  pages={203--209},
  year={2002},
  doi={10.1016/s1535-6108(02)00030-2},
  publisher={Elsevier}
}

@article{method_glmnet,
  title = {Regularization Paths for Generalized Linear Models via Coordinate Descent},
  author = {Jerome Friedman and Trevor Hastie and Robert Tibshirani},
  journal = {Journal of Statistical Software},
  year = {2010},
  volume = {33},
  number = {1},
  pages = {1--22},
  doi = {10.18637/jss.v033.i01},
}

@Manual{method_BoomSpikeSlab,
  title = {BoomSpikeSlab: MCMC for Spike and Slab Regression},
  author = {Steven L. Scott},
  year = {2023},
  note = {R package version 1.2.6},
  url = {https://CRAN.R-project.org/package=BoomSpikeSlab},
  doi = {10.32614/CRAN.package.BoomSpikeSlab},
}

@article{method_islasso,
  title = {The Induced Smoothed lasso: A practical framework for hypothesis testing in high dimensional regression},
  author = {Giovanna Cilluffo and Gianluca Sottile and Stefania {La Grutta} and {{V.M.R. Muggeo}}},
  journal = {Statistical Methods in Medical Research},
  year = {2020},
  volume = {29},
  number = {3},
  pages = {765--777},
  doi = {10.1177/0962280219842890},
}

@article{bayarri2012criteria,
    author = {M. J. Bayarri and J. O. Berger and A. Forte and G. Garc{\'i}a-Donato},
    title = {{Criteria for {B}ayesian model choice with application to variable selection}},
    volume = {40},
    journal = {Annals of Statistics},
    number = {3},
    publisher = {Institute of Mathematical Statistics},
    pages = {1550 -- 1577},
    year = {2012},
    doi = {10.1214/12-AOS1013},
    URL = {https://doi.org/10.1214/12-AOS1013}
}

@article{Brandt-2021,
  title = {Political Psychology Data from a 26-wave Yearlong Longitudinal Study (2019–2020)},
  volume = {9},
  ISSN = {2050-9863},
  url = {http://dx.doi.org/10.5334/jopd.54},
  DOI = {10.5334/jopd.54},
  number = {1},
  journal = {Journal of Open Psychology Data},
  publisher = {Ubiquity Press,  Ltd.},
  author = {Brandt,  Mark J. and Turner-Zwinkels,  Felicity M. and Kubin,  Emily},
  year = {2021},
  month = jul,
  pages = {2}
}

@article{Zhang2019,
  title={A novel variational {B}ayesian method for variable selection in logistic regression models},
  author={Zhang, Chun-Xia and Xu, Shuang and Zhang, Jiang-She},
  journal={Computational Statistics \& Data Analysis},
  volume={133},
  pages={1--19},
  year={2019}
}

@article{Shi2019,
  title={Bayesian model selection for generalized linear models using non-local priors},
  author={Shi, Guiling and Lim, Chae Young and Maiti, Tapabrata},
  journal={Computational Statistics \& Data Analysis},
  volume={133},
  pages={285--296},
  year={2019}
}

@article{Fouskakis2018,
  title={Power-expected-posterior priors for generalized linear models},
  author={Fouskakis, Dimitris and Ntzoufras, Ioannis and Perrakis, Konstantinos},
  year={2018},
 journal = "Bayesian Analysis",
 Volume = 13,
 pages = "721--748"
}

@article{Griffin2020,
  title={Modelling environmental {DNA} data; {B}ayesian variable selection accounting for false positive and false negative errors},
  author={Griffin, Jim E and Matechou, Eleni and Buxton, Andrew S and Bormpoudakis, Dimitrios and Griffiths, Richard A},
  journal={Journal of the Royal Statistical Society Series C: Applied Statistics},
  volume={69},
  number={2},
  pages={377--392},
  year={2020}
}

@article{wang2004comparison,
  title={Comparison of {B}ayesian model averaging and stepwise methods for model selection in logistic regression},
  author={Wang, Duolao and Zhang, Wenyang and Bakhai, Ameet},
  journal={Statistics in Medicine},
  volume={23},
  pages={3451--3467},
  year={2004},
  doi={10.1002/sim.1930},
  publisher={Wiley}
}

@article{Berkson1944,
  author    = {Joseph Berkson},
  title     = {Application of the logistic function to bio-assay},
  journal   = {Journal of the American Statistical Association},
  volume    = {39},
  number    = {227},
  pages     = {357--365},
  year      = {1944},
  publisher = {Taylor \& Francis},
  doi       = {10.1080/01621459.1944.10500699}
}

@article{Cox1958,
  author  = {D. R. Cox},
  title   = {The regression analysis of binary sequences (with discussion)},
  journal = {Journal of the Royal Statistical Society: Series B (Methodological)},
  volume  = {20},
  number  = {2},
  pages   = {215--242},
  year    = {1958}
}

@book{HosmerLemeshow2013,
  author    = {David W. Hosmer and Stanley Lemeshow and Rodney X. Sturdivant},
  title     = {Applied Logistic Regression},
  publisher = {Wiley},
  edition   = {3rd},
  year      = {2013},
  series    = {Wiley Series in Probability and Statistics},
  isbn      = {978-0470582473}
}

@article{Tuchler01032008,
author = {Regina T\"{u}chler},
title = {Bayesian Variable Selection for Logistic Models Using Auxiliary Mixture Sampling},
journal = {Journal of Computational and Graphical Statistics},
volume = {17},
number = {1},
pages = {76--94},
year = {2008},
publisher = {ASA Website},
doi = {10.1198/106186008X289849},
URL = {https://doi.org/10.1198/106186008X289849}
}

@article{ChenEtAl_1999,
 author = {Ming-Hui Chen and Joseph G. Ibrahim and Constantin Yiannoutsos},
journal = {{J. R. Stat. Soc. Ser. B (Stat. Methodol.)}},
 number = {1},
 pages = {223--242},
 title = {Prior Elicitation, Variable Selection and {B}ayesian Computation for Logistic Regression Models},
 volume = {61},
 year = {1999}
}

@article{Wagner_2012,
title = {Bayesian model selection for logistic regression models with random intercept},
journal = {Computational Statistics \& Data Analysis},
volume = {56},
number = {5},
pages = {1256-1274},
year = {2012},
doi = {https://doi.org/10.1016/j.csda.2011.06.033},
author = {Helga Wagner and Christine Duller}
}

@article{Raftery_1996_approx,
    author = {Raftery, Adrian E.},
    title = {Approximate {B}ayes factors and accounting for model uncertainty in generalised linear models},
    journal = {Biometrika},
    volume = {83},
    number = {2},
    pages = {251-266},
    year = {1996},
    doi = {10.1093/biomet/83.2.251},
    url = {https://doi.org/10.1093/biomet/83.2.251}
}

@article{Porwal_2024,
author = {Anupreet Porwal and Abel Rodr{\'i}guez},
title = {{Laplace power-expected-posterior priors for logistic regression}},
volume = {19},
journal = {Bayesian Analysis},
number = {4},
pages = {1163 -- 1186},
year = {2024},
doi = {10.1214/23-BA1389},
}

@article{PorwalRaftery_2022,
author = {Porwal, A. and Raftery, A. E.},
title = {Comparing methods for statistical inference with model uncertainty},
journal = {Proceedings of the National Academy of Sciences},
volume = {119},
number = {16},
pages = {e2120737119},
year = {2022},
doi = {10.1073/pnas.2120737119}
}

@article{Derksen1992,
  author    = {Shelley Derksen and H. J. Keselman},
  title     = {Backward, forward and stepwise automated subset selection algorithms: Frequency of obtaining authentic and noise variables},
  journal   = {British Journal of Mathematical and Statistical Psychology},
  volume    = {45},
  number    = {2},
  pages     = {265--282},
  year      = {1992},
  doi       = {10.1111/j.2044-8317.1992.tb00992.x}
}

@article{Austin2004,
  author    = {Peter C. Austin and Jack V. Tu},
  title     = {Automated variable selection methods for logistic regression produced unstable models for predicting acute myocardial infarction mortality},
  journal   = {Journal of Clinical Epidemiology},
  volume    = {57},
  number    = {11},
  pages     = {1138--1146},
  year      = {2004},
  doi       = {10.1016/j.jclinepi.2004.04.003}
}

@article{tibshirani1996regression,
  title={Regression shrinkage and selection via the lasso},
  author={Tibshirani, Robert},
  journal={Journal of the Royal Statistical Society: Series B (Methodological)},
  volume={58},
  number={1},
  pages={267--288},
  year={1996}
}

@book{buhlmann2011statistics,
  title={Statistics for High-Dimensional Data: Methods, Theory and Applications},
author = {B\"{u}hlmann, Peter and van de Geer, Sara},
  year={2011},
  publisher={Springer},
  address={Heidelberg}
}

@article{park2007l1,
  title={An {L}1-regularization path algorithm for generalized linear models},
  author={Park, Myunghee Cho and Hastie, Trevor},
journal = {{J. R. Stat. Soc. Ser. B (Stat. Methodol.)}},
  volume={69},
  number={4},
  pages={659--677},
  year={2007}
}

@article{hoerl1970ridge,
  title={Ridge regression: Biased estimation for nonorthogonal problems},
  author={Hoerl, Arthur E and Kennard, Robert W},
  journal={Technometrics},
  volume={12},
  number={1},
  pages={55--67},
  year={1970}
}

@article{zou2005regularization,
  title={Regularization and variable selection via the elastic net},
  author={Zou, Hui and Hastie, Trevor},
  journal={Journal of the Royal Statistical Society: Series B (Statistical Methodology)},
  volume={67},
  number={2},
  pages={301--320},
  year={2005}
}

@article{van2014asymptotically,
  title={On asymptotically optimal confidence regions and tests for high-dimensional models},
  author={van de Geer, Sara and B\"{u}hlmann, Peter and Ritov, Ya'acov and Dezeure, Ruben},
  journal={Annals of Statistics},
  volume={42},
  number={3},
  pages={1166--1202},
  year={2014}
}

@article{javanmard2014confidence,
  title={Confidence intervals and hypothesis testing for high-dimensional regression},
  author={Javanmard, Adel and Montanari, Andrea},
  journal={Journal of Machine Learning Research},
  volume={15},
  number={1},
  pages={2869--2909},
  year={2014}
}

@article{rockova2018spike,
  title={The spike-and-slab lasso},
  author={Ro{\v{c}}kov{\'a}, Veronika and George, Edward I},
  journal={Journal of the American Statistical Association},
  volume={113},
  number={521},
  pages={431--444},
  year={2018}
}

@article{breheny2011coordinate,
  title={Coordinate descent algorithms for nonconvex penalized regression, with applications to biological feature selection},
  author={Breheny, Patrick and Huang, Jian},
  journal={Annals of Applied Statistics},
  volume={5},
  number={1},
  pages={232--253},
  year={2011},
  doi={10.1214/10-AOAS388},
  publisher={Institute of Mathematical Statistics}
}

@article{lecessie1992ridge,
  title={Ridge estimators in logistic regression},
  author={Le Cessie, S and Van Houwelingen, J C},
  journal={Applied Statistics},
  volume={41},
  number={1},
  pages={191--201},
  year={1992},
  doi={10.2307/2347628}
}

@article{firth1993bias,
  title={Bias reduction of maximum likelihood estimates},
  author={Firth, David},
  journal={Biometrika},
  volume={80},
  number={1},
  pages={27--38},
  year={1993},
  doi={10.1093/biomet/80.1.27},
  publisher={Oxford University Press}
}

@article{heinze2002solution,
  title={A solution to the problem of separation in logistic regression},
  author={Heinze, Georg and Schemper, Michael},
  journal={Statistics in Medicine},
  volume={21},
  number={16},
  pages={2409--2419},
  year={2002},
  doi={10.1002/sim.1047},
  publisher={Wiley}
}

@article{albert1984existence,
  title={On the existence of maximum likelihood estimates in logistic regression models},
  author={Albert, Arthur and Anderson, John A},
  journal={Biometrika},
  volume={71},
  number={1},
  pages={1--10},
  year={1984},
  doi={10.1093/biomet/71.1.1}
}

@article{heinze2006comparative,
  title={A comparative investigation of methods for logistic regression with separated or nearly separated data},
  author={Heinze, Georg},
  journal={Statistics in Medicine},
  volume={25},
  number={24},
  pages={4216--4226},
  year={2006},
  doi={10.1002/sim.2687},
  publisher={Wiley}
}

@article{kotani2025goodness,
  title={A goodness-of-fit measure for logistic regression under separation},
  author={Kotani, Naoki and Kurosawa, Takeshi and Eshima, Nobuoki},
  journal={Communications in Statistics - Theory and Methods},
  volume={54},
  number={13},
  pages={4083--4100},
  year={2025},
  doi={10.1080/03610926.2024.2413845},
  publisher={Taylor \& Francis}
}

@article{akaike1974new,
  author  = {Akaike, Hirotugu},
  title   = {A New Look at the Statistical Model Identification},
  journal = {IEEE Transactions on Automatic Control},
  volume  = {19},
  number  = {6},
  pages   = {716--723},
  year    = {1974}
}

@book{burnham2002model,
  title={Model Selection and Multimodel Inference: {A} Practical Information-theoretic Approach},
  author={Burnham, Kenneth P and Anderson, David R},
  year={2002},
  publisher={Springer}
}

@article{gelman2008prior,
  title={A weakly informative default prior distribution for logistic and other regression models},
  author={Gelman, Andrew and Jakulin, Aleks and Pittau, Maria Grazia and Su, Yu-Sung},
  journal={Annals of Applied Statistics},
  volume={2},
  number={4},
  pages={1360--1383},
  year={2008},
  doi={10.1214/08-AOAS191}
}

@book{miller2002subset,
  title={Subset Selection in Regression},
  author={Miller, Alan},
  year={2002},
  publisher={CRC Press}
}

@article{Freedman1983,
author = "D. A. Freedman",
year = 1983,
title = "A note on screening regression equations",
journal = "American Statistician",
volume = 37,
pages = "152--155"
}

@book{Leamer1978,
  title={Specification Searches: Ad hoc Inference with Nonexperimental Data},
  author={Leamer, Edward E},
  volume={53},
  year={1978},
  publisher={Wiley}
}

@techreport{Raftery1988,
author = "Raftery, Adrian E",
year = 1988,
title = "Approximate {B}ayes factors for generalized linear models",
type = "Technical Report",
number = 121,
institution = "Department of Statistics, University of Washington",
note = "https://stat.uw.edu/sites/default/files/files/reports/1988/tr121.pdf"
}

@article{George1993,
  title={Variable selection via {G}ibbs sampling},
  author={George, Edward I and McCulloch, Robert E},
  journal={Journal of the American Statistical Association},
  volume={88},
  number={423},
  pages={881--889},
  year={1993},
  publisher={Taylor \& Francis}
}

@article{Madigan1994,
  title={Model selection and accounting for model uncertainty in graphical models using {O}ccam's window},
  author={Madigan, David and Raftery, Adrian E},
  journal={Journal of the American Statistical Association},
  volume={89},
  number={428},
  pages={1535--1546},
  year={1994},
  publisher={Taylor \& Francis}
}

@article{kass1995bayes,
  title={Bayes factors},
  author={Kass, Robert E and Raftery, Adrian E},
  journal={Journal of the American Statistical Association},
  volume={90},
  number={430},
  pages={773--795},
  year={1995},
}

@article{fernandez2001benchmark,
  title={Benchmark priors for {B}ayesian model averaging},
  author={Fern\'andez, Carmen and Ley, Eduardo and Steel, Mark F. J.},
  journal={Journal of Econometrics},
  volume={100},
  number={2},
  pages={381--427},
  year={2001},
  publisher={Elsevier}
}

@article{Wasserman2000,
  title={Bayesian model selection and model averaging},
  author={Wasserman, Larry},
  journal={Journal of Mathematical Psychology},
  volume={44},
  number={1},
  pages={92--107},
  year={2000},
  publisher={Elsevier}
}

@article{Hoeting1999,
  title={Bayesian model averaging: {A} tutorial},
  author={Hoeting, Jennifer A and Madigan, David and Raftery, Adrian E and Volinsky, Chris T},
  journal={Statistical Science},
  volume={14},
  pages={382--417},
  year={1999},
}

@article{Clyde2004,
  title={Model uncertainty},
  author={Clyde, Merlise and George, Edward I},
  journal={Statistical Science},
  volume={19},
  number={1},
  pages={81--94},
  year={2004},
  publisher={Institute of Mathematical Statistics}
}

@article{Fragoso2018,
  title={Bayesian model averaging: A systematic review and conceptual classification},
  author={Fragoso, Tiago M and Bertoli, Wesley and Louzada, Francisco},
  journal={International Statistical Review},
  volume={86},
  number={1},
  pages={1--28},
  year={2018},
  publisher={Wiley Online Library}
}

@article{Forte2018,
  title={Methods and tools for {B}ayesian variable selection and model averaging in normal linear regression},
  author={Forte, Anabel and Garcia-Donato, Gonzalo and Steel, Mark F. J.},
  journal={International Statistical Review},
  volume={86},
  number={2},
  pages={237--258},
  year={2018},
  publisher={Wiley Online Library}
}

@article{Kaplan_2021,
	author = {Kaplan, D.},
	doi = {10.1007/s11336-021-09754-5},
	journal = {Psychometrika},
	number = {1},
	pages = {215--238},
	title = {On the quantification of model uncertainty: {A} {B}ayesian perspective},
	volume = {86},
	year = {2021}}

@article{Rockova2014,
  title={{EMVS}: The {EM} approach to {B}ayesian variable selection},
  author={Ro{\v{c}}kov{\'a}, Veronika and George, Edward I},
  journal={Journal of the American Statistical Association},
  volume={109},
  number={506},
  pages={828--846},
  year={2014},
  publisher={Taylor \& Francis}
}

@article{womack2014inference,
  title={Inference from intrinsic {B}ayes’ procedures under model selection and uncertainty},
  author={Womack, Andrew J and Le{\'o}n-Novelo, Luis and Casella, George},
  journal={Journal of the American Statistical Association},
  volume={109},
  number={507},
  pages={1040--1053},
  year={2014},
  publisher={Taylor \& Francis}
}

@article{fouskakis2015power,
  title={Power-expected-posterior priors for variable selection in {G}aussian linear models},
  author={Fouskakis, Dimitris and Ntzoufras, Ioannis and Draper, David},
  journal={Bayesian Analysis},
  volume={10},
  number={1},
  pages={75--107},
  year={2015},
  publisher={International Society for Bayesian Analysis}
}

@article{Park2008,
  title={The {B}ayesian lasso},
  author={Park, Trevor and Casella, George},
  journal={Journal of the American Statistical Association},
  volume={103},
  number={482},
  pages={681--686},
  year={2008},
  publisher={Taylor \& Francis}
}

@article{zhang2010nearly,
  title = {Nearly unbiased variable selection under minimax concave penalty},
  volume = {38},
  ISSN = {0090-5364},
  url = {http://dx.doi.org/10.1214/09-aos729},
  DOI = {10.1214/09-aos729},
  number = {2},
  journal = {Annals of Statistics},
  publisher = {Institute of Mathematical Statistics},
  author = {Zhang,  Cun-Hui},
  year = {2010},
  pages = {894--942}
}

@article{fan2001variable,
  title={Variable selection via nonconcave penalized likelihood and its oracle properties},
  author={Fan, Jianqing and Li, Runze},
  journal={Journal of the American Statistical Association},
  volume={96},
  number={456},
  pages={1348--1360},
  year={2001},
  publisher={Taylor \& Francis}
}

@article{Holmes2011discussion,
author = "C. C. Holmes",
year = 2011,
title = "Discussion of `{R}egression shrinkage and selection via the lasso:
{A} retrospective'",
journal = "Journal of the Royal Statistical Society, Series B",
volume = 73,
pages = "279--280"
}

@article{kyung2010penalized,
  title={Penalized regression, standard errors, and {B}ayesian lassos},
  author={Casella, George and Ghosh, Malay and Gill, Jeff and Kyung, Minjung},
  journal={Bayesian Analysis},
  volume={5},
  number={2},
  pages={369--411},
  year={2010},
  publisher={International Society for Bayesian Analysis}
}

@article{brier1950verification,
  title={Verification of forecasts expressed in terms of probability},
  author={Brier, Glenn W},
  journal={Monthly Weather Review},
  volume={78},
  number={1},
  pages={1--3},
  year={1950}
}

@article{raftery1995bayesian,
  title={Bayesian model selection in social research},
  author={Raftery, Adrian E},
  journal={Sociological Methodology},
  volume={25},
  pages={111--163},
  year={1995},
  publisher={JSTOR}
}

@article{young2014fast,
  title={Fast {B}ayesian inference for gene regulatory networks using ScanBMA},
  author={Young, William Chad and Raftery, Adrian E and Yeung, Ka Yee},
  journal={BMC Systems Biology},
  volume={8},
  number={1},
  pages={47},
  year={2014},
  publisher={BioMed Central}
}

@article{liang2008mixtures,
  title={Mixtures of g-priors for {B}ayesian variable selection},
  author={Liang, Feng and Paulo, Rui and Molina, German and Clyde, Merlise A and Berger, Jim O},
  journal={Journal of the American Statistical Association},
  volume={103},
  pages={410--423},
  year={2008},
  publisher={Taylor \& Francis}
}

@article{george2000calibration,
  title={Calibration and empirical {B}ayes variable selection},
  author={George, Edward I and Foster, Dean P},
  journal={Biometrika},
  volume={87},
  number={4},
  pages={731--747},
  year={2000},
  publisher={Oxford University Press}
}

@article{clyde2000flexible,
  title={Flexible empirical {B}ayes estimation for wavelets},
  author={Clyde, Merlise and George, Edward I},
  journal={Journal of the Royal Statistical Society: Series B, Statistical Methodology},
  volume={62},
  number={4},
  pages={681--698},
  year={2000},
}

@incollection{zellner1986assessing,
  author    = {Arnold Zellner},
  title     = {On Assessing Prior Distributions and {B}ayesian Regression Analysis with g-Prior Distributions},
  booktitle = {Bayesian Inference and Decision Techniques: Essays in Honor of Bruno de Finetti},
  editor    = {Prem K. Goel and Arnold Zellner},
  publisher = {Elsevier Science Publishers},
  address   = {New York},
  pages     = {233--243},
  year      = {1986}
}

@misc{Berger2021,
    author       = {Berger, James O.},
    title        = {Four Types of Frequentism and Their Interplay with {B}ayesianism},
    year         = {2021},
    howpublished = {De Finetti Lecture, Meeting of the International Society for Bayesian Analysis}
}

@article{park2010estimation,
  title={Estimation of effect size distribution from genome-wide association studies and implications for future discoveries},
  author={Park, Ju-Hyun and Wacholder, Sholom and Gail, Mitchell H and Peters, Ulrike and Jacobs, Kevin B and Chanock, Stephen J and Chatterjee, Nilanjan},
  journal={Nature Genetics},
  volume={42},
  pages={570--575},
  year={2010},
}

@book{Jeffreys1961,
author = "H. Jeffreys",
year = 1961,
title = "Theory of Probability",
edition = "3rd",
publisher = "Oxford University Press",
address = "Oxford, U.K."
}

@misc{Mattei2020,
  author       = {Mattei, P. A.},
  title        = {A Parsimonious Tour of {B}ayesian Model Uncertainty},
  year         = {2020},
  note         = {arXiv:1902.05539},
  howpublished = {\url{https://arxiv.org/abs/1902.05539}}
}

@article{RubinSchenker1986,
  title={Efficiently simulating the coverage properties of interval estimates},
  author={Rubin, Donald B and Schenker, Nathaniel},
  journal={Journal of the Royal Statistical Society: Series C, Applied Statistics},
  volume={35},
  number={2},
  pages={159--167},
  year={1986},
}

@article{raftery2003discussion,
  title={Discussion: Performance of {B}ayesian model averaging},
  author={Raftery, Adrian E and Zheng, Yingye},
  journal={Journal of the American Statistical Association},
  volume={98},
  number={464},
  pages={931--938},
  year={2003},
  publisher={Taylor \& Francis}
}

@article{hans2009bayesian,
  title={Bayesian lasso regression},
  author={Hans, Chris},
  journal={Biometrika},
  volume={96},
  number={4},
  pages={835--845},
  year={2009},
  publisher={Oxford University Press}
}

@manual{R-aplore3,
  title        = {aplore3: Applied Logistic Regression (3rd Edition) Data Sets},
  author       = {Bret Larget},
  year         = {2017},
  note         = {R package version 0.1},
  doi          = {10.32614/CRAN.package.aplore3},
  url          = {https://CRAN.R-project.org/package=aplore3},
}

@misc{mitgandhi2023logistic,
  author       = {Gandhi, Mit},
  title        = {Dataset for Logistic Regression},
  year         = {2023},
  howpublished = {\url{https://www.kaggle.com/datasets/mitgandhi10/dataset-for-logistic-regression}},
  note         = {Kaggle dataset}
}

@article{Dbska2021,
  title = {The cognitive basis of dyslexia in school‐aged children: A multiple case study in a transparent orthography},
  volume = {25},
  ISSN = {1467-7687},
  url = {http://dx.doi.org/10.1111/desc.13173},
  DOI = {10.1111/desc.13173},
  number = {2},
  journal = {Developmental Science},
  publisher = {Wiley},
  author = {D\k{e}bska,  Agnieszka and {\L}uniewska, Magdalena and Zubek, Julian and Chyl, Katarzyna and Dynak, Agnieszka and Dzi\k{e}giel‐Fivet, Gabriela and Plewko, Joanna and Jednor\'{o}g, Katarzyna and Grabowska, Anna},
  year = {2021}
}

@misc{osfDbska2021,
  author       = {Dębska, Agnieszka and Łuniewska, Magdalena},
  title        = {Cognitive basis of dyslexia in school-aged children: {A} multiple case study in a transparent orthography},
  year         = {2020},
  note         = {Open Science Framework},
  howpublished = {\url{https://osf.io/mj32v}},
}

@misc{yasserh2023housing,
  author       = {{Yasser H}},
  title        = {Housing Prices Dataset},
  year         = {2023},
  howpublished = {\url{https://www.kaggle.com/datasets/yasserh/housing-prices-dataset/data}},
  note         = {Kaggle dataset}
}

@misc{johnsmith88_heart_disease,
  author       = {David Lapp},
  title        = {Heart Disease Dataset},
  year         = {2023},
  note         = {Kaggle dataset},
  howpublished = {\url{https://www.kaggle.com/datasets/johnsmith88/heart-disease-dataset}},
}

@misc{UCIdivorce,
  author       = {{UCI Machine Learning Repository}},
  title        = {Divorce Predictors data set},
  year         = {2019},
  howpublished = {\url{https://doi.org/10.24432/C53W5P}},
  note         = {UCI Machine Learning Repository}
}

@manual{nki70-R-penalized,
  title   = {penalized: {L1} (Lasso and Fused Lasso) and {L2} (Ridge) Penalized Estimation in {GLMs} and in the {C}ox Model},
  author  = {Jelle J. Goeman and Rosa J. Meijer and Nimisha Chaturvedi and Matthew Lueder},
  year    = {2025},
  note    = {R package version 0.9-53},
  url     = {https://CRAN.R-project.org/package=penalized}
}

@misc{chapman1994musk,
  author       = {Chapman, D. and Jain, A.},
  title        = {Musk (Version 2)},
  year         = {1994},
  howpublished = {https://doi.org/10.24432/C51608},
  note = {UCI Machine Learning Repository}
}

@article{osf-covid19,
    title     = {Social cohesion predicts {COVID}-19 vaccination intentions and uptake},
    volume    = {17},
    number    = {7},
    journal   = {Social and Personality Psychology Compass},
    author    = {C{\'a}rdenas, Diana and Orazani, Nima and Manueli, Farah and Donaldson, Jessica L. and Stevens, Mark and Cruwys, Tegan and Platow, Michael J. and O'Donnell, James and Zekulin, Michael G. and Qureshi, Israr and Walker, Iain and Reynolds, Katherine J.},
    year      = {2023},
    pages     = {e12759},
    doi       = {10.1111/spc3.12759}
}

@article{WuEtAl2016,
author = {Wu, H.-H. and Ferreira, M. A. R. and Gompper, M. E.},
title = {{Consistency of hyper-$g$-prior-based {B}ayesian variable selection for generalized linear models}},
volume = {30},
journal = {Brazilian Journal of Probability and Statistics},
number = {4},
publisher = {Brazilian Statistical Association},
pages = {691 -- 709},
year = {2016},
doi = {10.1214/15-BJPS299}
}

@article{Gneiting2007,
  title={Strictly proper scoring rules, prediction, and estimation},
  author={Gneiting, Tilmann and Raftery, Adrian E},
  journal={Journal of the American Statistical Association},
  volume={102},
  pages={359--378},
  year={2007}
}

@book{ClaeskensHjort2008,
  title={Model Selection and Model Averaging},
  author={Claeskens, Gerda and Hjort, Nils Lid},
  journal={Cambridge books},
  year={2008},
  publisher={Cambridge University Press}
}

@article{HjortClaeskens2003,
  title={Frequentist model average estimators},
  author={Hjort, Nils Lid and Claeskens, Gerda},
  journal={Journal of the American Statistical Association},
  volume={98},
  number={464},
  pages={879--899},
  year={2003}
}

@article{Steel2020,
  title={Model averaging and its use in economics},
  author={Steel, Mark FJ},
  journal={Journal of Economic Literature},
  volume={58},
  number={3},
  pages={644--719},
  year={2020}
}

@book{BernardoSmith1994,
  title={Bayesian Theory},
  author={Bernardo, Jos{\'e} M and Smith, Adrian F M},
  year={1994},
  publisher={Wiley Online Library}
}

@article{SabanesBoveHeld2011,
    author = {Saban\'{e}s Bov\'{e}, D. and Held, L.},
    year = {2011},
    title = {Hyper-$g$ priors for generalized linear models},
    journal = {Bayesian Analysis},
    volume = {6},
    issue = {3},
    pages = {387--410},
    doi = {10.1214/11-BA615}
}

@article{HuJohnson2009,
    author = {Hu, J. and Johnson, V. E.},
    year = {2009},
    title = {Bayesian model selection using test statistics},
    journal = {Journal of the Royal Statistical Society Series B: Statistical Methodology},
    volume = {71},
    issue = {1},
    pages = {143--158},
    doi = {10.1111/j.1467-9868.2008.00678.x}
}

@article{HeldSabanesBoveGravestock2015,
    author = {Held, L. and Saban\'{e}s Bov\'{e}, D. and Gravestock, I.},
    year = {2015},
    title = {Approximate {B}ayesian model selection with the deviance statistic},
    journal = {Statistical Science},
    volume = {30},
    issue = {2},
    pages = {242--257},
    doi = {10.1214/14-STS510}
}

@article{GhoshLiMitra2018,
    author = {Ghosh, J. and Li, Y. and Mitra, R.},
    year = {2018},
    title = {On the use of {C}auchy prior distributions for {B}ayesian logistic regression},
    journal = {Bayesian Analysis},
    volume = {13},
    issue = {2},
    pages = {359--383},
    doi = {10.1214/17-BA1051}
}

@article{Ghosh2019,
    author = {Ghosh, J.},
    year = {2019},
    title = {{C}auchy and other shrinkage priors for logistic regression in the presence of separation},
    journal = {WIREs Computational Statistics},
    volume = {11},
    issue = {6},
    pages = {e1478},
    doi = {10.1002/wics.1478}
}

@article{LeySteel2012,
  title={Mixtures of $g$-priors for {B}ayesian model averaging with economic applications},
  author={Ley, E. and Steel, M. F. J.},
  journal={Journal of Econometrics},
  volume={171},
  number={2},
  pages={251-266},
  year={2012},
  doi={10.1016/j.jeconom.2012.06.009}
}

@article{Strawderman1971,
  title={Proper {B}ayes minimax estimators of the multivariate normal mean},
  author={Strawderman, W. E.},
  journal={Annals of Mathematical Statistics},
  volume={42},
  number={1},
  pages={385--388},
  year={1971}
}

@article{greenland2016penalization,
  title={Penalization, bias reduction, and default priors in logistic and related categorical and survival regressions},
  author={Greenland, S. and Mansournia, M. A.},
  journal={Statistics in Medicine},
  volume={34},
  number={23},
  pages={3133--3143},
  year={2015},
  doi={10.1002/sim.6537}
}

\end{document}